\documentclass[a4paper]{article}

\usepackage{INTERSPEECH2020}

\title{Cross-Lingual Speaker Verification with Domain-Balanced Hard Prototype Mining and Language-Dependent Score Normalization}
\name{Jenthe Thienpondt, Brecht Desplanques, Kris Demuynck}

\address{
    IDLab, Department of Electronics and Information Systems, Ghent University - imec, Belgium
}
\email{jenthe.thienpondt@ugent.be, brecht.desplanques@ugent.be}

\begin{document}

\maketitle
\begin{abstract}
In this paper we describe the top-scoring IDLab submission for the text-independent task of the Short-duration Speaker Verification (SdSV) Challenge 2020. The main difficulty of the challenge exists in the large degree of varying phonetic overlap between the potentially cross-lingual trials, along with the limited availability of in-domain DeepMine Farsi training data. We introduce domain-balanced hard prototype mining to fine-tune the state-of-the-art ECAPA-TDNN x-vector based speaker embedding extractor. The sample mining technique efficiently exploits speaker distances between the speaker prototypes of the popular AAM-softmax loss function to construct challenging training batches that are balanced on the domain-level. To enhance the scoring of cross-lingual trials, we propose a language-dependent s-norm score normalization. The imposter cohort only contains data from the Farsi target-domain which simulates the enrollment data always being Farsi. In case a Gaussian-Backend language model detects the test speaker embedding to contain English, a cross-language compensation offset determined on the AAM-softmax speaker prototypes is subtracted from the maximum expected imposter mean score. A fusion of five systems with minor topological tweaks resulted in a final MinDCF and EER of 0.065 and 1.45\% respectively on the SdSVC evaluation set.

\end{abstract}

\section{Introduction}
Speaker verification systems have improved significantly by the strength of deep learning \cite{x_vectors, x_vector_wide} and the increase in publicly available labeled training data \cite{vox1, vox2}.
However, most of these datasets tend to focus on the Anglosphere, making it hard to produce speaker embeddings that perform well on out-of-domain data. 

The SdSV Challenge uses this notion to create a challenging set of speaker verification trials, divided in two separate tasks. Task 1 consists of text-dependent speaker verification, for which both the lexical content and speaker identity should be equal across the enrollment and test utterances to indicate a valid trial. Task 2 is concerned with text-independent speaker verification, which only takes the speaker identities into account. This paper focuses solely on our submission to the second text-independent task.

Task 2 systems can only use a fixed training dataset consisting of VoxCeleb1 \cite{vox1}, VoxCeleb2 \cite{vox2}, LibriSpeech \cite{libri} and a part of the DeepMine corpus \cite{deepmine} containing in-domain Farsi training utterances across 588 speakers. Trials consists of producing a speaker similarity score between multiple Farsi enrollment utterances and a test utterance. The test utterance can either contain Farsi or English speech. Consequently, speaker verification systems should be able to reduce the language bias in cross-lingual trials. More details about the SdSV Challenge conditions can be found in the evaluation plan \cite{sdsv_plan}.



The rest of the paper is organized as follows: Section~\ref{sec:baseline} will describe the IDLab SdSVC final submission. The state-of-the-art ECAPA-TDNN~\cite{ecapa_tdnn} architecture is combined with adapted training procedures and backend scoring to tackle the challenge-specific difficulties. It is followed by a more in-depth analysis of the proposed approach in Section~\ref{sec:exp}. Section~\ref{sec:conclusion} will give the concluding remarks.

\section{SdSVC IDLab submission}

\label{sec:baseline}

This section is a system description of the IDLab SdSVC final submission. We start with a single system ECAPA-TDNN baseline~\cite{ecapa_tdnn}. The subsequent sections will tackle the problems of domain adaptation and cross-lingual language effects present in the SdSV Challenge data. The final subsection discusses system fusion.

\subsection{The ECAPA-TDNN baseline system}

\begin{figure}[t]
\centering
\includegraphics[scale=0.28]{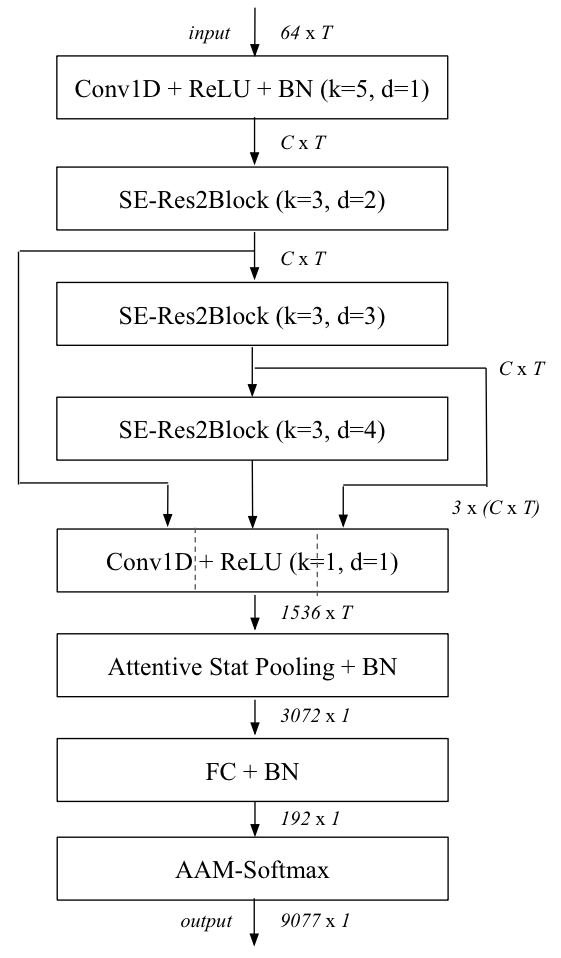}
\caption{{Network topology of the ECAPA-TDNN. We denote $k$ for kernel size and $d$ for dilation spacing of the Conv1D layers or SE-Res2Blocks. $C$ and $T$ correspond to the channel and temporal dimension of the intermediate feature-maps respectively.}}
\label{fig:full_network}
\end{figure}

All submitted speaker verification systems make use of the ECAPA-TDNN architecture proposed in~\cite{ecapa_tdnn}. This architecture is based on the well-known x-vector topology \cite{x_vectors} and introduces several enhancements to extract more robust speaker embeddings. It incorporates Squeeze-Excitation (SE) blocks~\cite{se_block}, multi-scale Res2Net~\cite{res2net} features, multi-layer feature aggregation~\cite{multi_stage_aggregation} and channel-dependent attentive statistics poolings~\cite{ecapa_tdnn}. The network topology is shown in Figure~\ref{fig:full_network}. Implementation details and performance analysis of this architecture can be found in~\cite{ecapa_tdnn}. We deviate slightly from the original architecture by also incorporating SE-Blocks in the residual connections.

We use all allowed training data, except the VoxCeleb1 test partition and LibriSpeech, for which only the \textit{train-other-500} subset~\cite{libri} is considered. This amounts to 9077 training speakers. We create 9 additional augmented copies of the training data following the Kaldi recipe \cite{x_vector_wide} in combination with the MUSAN corpus (babble, noise, music) \cite{musan} and the RIR\cite{rirs} dataset (reverb). The remaining augmentations are generated with the open-source SoX (tempo up, tempo down, phaser and flanger) and FFmpeg (alternating opus and aac compression) libraries.

To avoid overfitting during the ECAPA-TDNN training process, we take a random crop of 2 to 3 seconds of the utterances during each iteration. Similarly, we incorporate SpecAugment \cite{specaugment} as an online augmentation method which randomly masks 0 to 5 time frames and 0 to 8 frequency bands of the training log mel-spectrograms. The input features are 64-dimensional MFCCs from a 25~ms window with a 10~ms frame shift. The MFCCs are normalized through cepstral mean subtraction and no voice activity detection is applied.


We use the Angular Additive Margin (AAM) softmax \cite{arcface} as training criterion for the model. The system is trained with the Adam optimizer \cite{adam} until convergence on a small SdSVC validation subset that contains about 2.5\% of the Farsi training utterances. The training protocol uses a cyclical learning rate schedule with the \textit{triangular2} policy~\cite{clr}. The learning rate is varied between a minimum of 1e-8 and decaying maximum of 1e-3 during cycles of 130k iterations. A weight decay of 2e-5 is applied on all weights of the model except for the AAM-softmax layer which uses a weight decay value of 2e-4. We use a mini-batch size of 128.

The speaker enrollment models are constructed by averaging the corresponding $L_2$-normalized enrollment embeddings produced by the final fully-connected layer of the ECAPA-TDNN. The verification trials are scored by calculating the cosine distance between the enrollment model and the test utterance embedding. Scores are normalized using top-40 adaptive s-normalization \cite{as_norm_original_1, as_norm_original_2}.  The imposter cohort consists of speakers represented by the average of all their length-normalized training embeddings. The final scores are calibrated with logistic regression \cite{bosaris} on our small SdSVC validation subset.







We consider five implementations with minor topological differences as shown in Table~\ref{tab:exp_results}. We alternate the embedding size between 192 and 256. The Res2Net multi scale features inside the SE-Res2Blocks are optionally replaced by the standard TDNN 1-dimensional dilated convolutions. \textit{Summed} indicates if the input of each SE-Res(2)Block is the sum of the output of all preceding SE-Res(2)Blocks instead of only considering the output of the preceding block. The number of filters in the convolutional frame layers $C$ is set to 1024, which is reduced to 512 in the bottleneck of the SE-Res(2)Blocks to limit the amount of model parameters. However, system 5 is developed without this constraint and the channel dimension is kept to 2048 for all feature maps in the frame layers.



\subsection{Hard prototype mining}
\label{sec:hpm}

To further improve performance on the baseline, we investigate how to exploit the information of the in-domain training data more efficiently. We combine targeting harder samples and putting more importance to target-domain samples with our proposed Hard Prototype Mining (HPM) fine-tuning strategy.

Hard negative mining in speaker recognition systems has mostly been explored in conjunction with metric learning objective functions \cite{triplet, proto, ge2e}. A current overview of these loss functions applied within speaker recognition is provided in \cite{metrics}. Metric learning objectives shift a lot of implementation challenges to the sample mining process. In contrast, HPM is a simple and computationally efficient hard negative mining method that interoperates with the AAM-softmax loss.

\begin{table*}
  \caption{EER and MinDCF performance of all individual systems and final fusion on the VoxCeleb1 and SdSVC 2020 test sets. All HPM models use our domain-balanced hard prototype mining technique as explained in Section \ref{sec:tdhpm}. LID denotes usage of our language-dependent s-normalization variant introduced in Section \ref{sec:ld_s_norm}.}
  \label{tab:exp_results}
  \centering
  \begin{tabular}{cccccccccc}
    \toprule
    \multicolumn{1}{c}{\textbf{\#}} & 
    \multicolumn{1}{c}{\textbf{System (\# params)}} &
    \multicolumn{1}{c}{\textbf{Emb. dim}} & 
    \multicolumn{1}{c}{\textbf{Res2}} &
    \multicolumn{1}{c}{\textbf{Summed}} &
    \multicolumn{1}{c}{\textbf{Fine-tune}} &
    \multicolumn{2}{c}{\textbf{VoxCeleb1}} &
    \multicolumn{2}{c}{\textbf{SdSVC 2020}} \\
    \cmidrule(lr){7-8} \cmidrule(lr){9-10}
    \multicolumn{6}{c}{\textbf{}} &
    \multicolumn{1}{c}{\textbf{EER(\%)}} & \multicolumn{1}{c}{\textbf{MinDCF}} &
    \multicolumn{1}{c}{\textbf{EER(\%)}} & \multicolumn{1}{c}{\textbf{MinDCF}} \\
    \midrule
     &  &  &  &  & baseline  & 0.94 & 0.1181 & 2.38 & 0.1042 \\
    1 & ECAPA-TDNN (24M)  & 192 & no & no & HPM & 0.85 & 0.0945 & 1.81 & 0.0798 \\
     &  & &  &  & HPM + LID  & - & -  & 1.75 & 0.0781 \\
    \midrule
     &  & &  &  & baseline  & 1.03 & 0.1260 & 2.34 & 0.0996 \\
    2 & ECAPA-TDNN (24M) & 192 & no & yes & HPM  & 0.96 & 0.1248 & 1.77 & 0.0791 \\
     &  & &  &  & HPM + LID  & - & -  & 1.72 & 0.0775 \\
    \midrule
     &  & &  &  & baseline  & 0.86 & 0.0969 & 2.32 & 0.1008 \\
    3 & ECAPA-TDNN (16M) & 256 & yes & no & HPM & 0.81 & 0.1033 & 1.75 & 0.0784 \\
     &  & &  &  & HPM + LID  & -  & -  & 1.69 & 0.0764 \\
    \midrule
     &  & &  &  & baseline  & 0.88 & 0.1101 & 2.32 & 0.0994 \\
    4 & ECAPA-TDNN (16M) & 256 & yes & yes & HPM & 0.88 & 0.1161 & 1.69 & 0.0759 \\
     &  & &  &  & HPM + LID  & -  & -  & 1.63 & 0.0742 \\
    \midrule
     &  & &  &  & baseline  & 0.87 & 0.0824 & 2.13 & 0.0938 \\
    5 & ECAPA-TDNN (44M) & 256 & yes & yes & HPM & 0.79 & 0.1010 & 1.69 & 0.0759 \\
     & & & & & HPM + LID  & - & - & \textbf{1.63} & \textbf{0.0739} \\
    \midrule
    \midrule
    & Weighted fusion of 1-5 & & & & HPM + LID & - & - & \textbf{1.45} &
    \textbf{0.0651} \\
    \bottomrule
  \end{tabular}
\end{table*}

\subsubsection{Broad hard prototype mining}
\label{sec:bhpm}


The general principle behind HPM is to detect hard speakers that confuse the speaker verification system the most and to subsequently construct batches with utterances from these speakers. A direct and continuous measurement of speaker confusion between all training samples would be computationally infeasible. Hence, we need an approximate and efficient way to compute training speaker similarities that can be easily updated as the training progresses.

We interpret the weights of the AAM-softmax layer as approximations of the class-centers of the training speakers and refer to them as speaker prototypes. As these trainable weights are already a part of the model, there are no additional computations needed.
Given batch size $n$ and $N$ training speakers, the AAM-softmax loss $L$ with margin $m$ is defined as:
\begin{equation}
\label{aam_softmax}
L = -\frac{1}{n} \sum^{n}_{i=1} log \frac{e^{s(cos(\theta_{y_{i}} +m ))}}{e^{s(cos(\theta_{y_{i}} +m))} + \sum_{j=1, j \neq y_i}^N e^{s(cos(\theta_{j}))}}
\end{equation}
where $\theta_{y_{i}}$ is the angle between the sample embedding $\pmb{x}_i$ with corresponding speaker identity $y_{i}$ and the speaker prototype $\pmb{W}_{y_{i}}$. $\theta_j$ is the angle with all other $L_2$-normalized speaker prototypes stored in a trainable matrix $\pmb{W} \in \mathbb{R}^{D \times N}$ with $D$ indicating the embedding size. A speaker similarity matrix $\pmb{S} \in \mathbb{R}^{N \times N}$ can be constructed from $\pmb{W}^T\pmb{W}$, containing the cosine distances between all pairs of speaker prototypes.

A straightforward way of constructing batches would be to only mine samples from the most difficult speaker pairs according to $\pmb{S}$. However, this could lead to oversampling a narrow group of speakers which potentially degrades generalization performance. Consequently, we construct mini-batches by iterating randomly over all $N$ training speakers. Each iteration determines $S$ speakers, irrespective of their similarity, for which $U$ random utterances are sampled from each of their $I$ most similar speakers, including the selected speaker. This implies that $S \times U \times I$ should be equal to the batch size $n$. When we have iterated over all training speakers, the similarity matrix $\pmb{S}$ is updated and the batch generating process is repeated. Experiments indicate that given a batch size of 128, $S=16$, $I=8$ and $U=1$ result in good performance.

To fine-tune all models in this paper, we reduce the maximum of the cyclical learning rate to 1e-4 and reduce the cycle length to 60k iterations. 

\subsubsection{Domain-balanced hard prototype mining}
\label{sec:tdhpm}

In the general HPM strategy discussed above, the $S$ selected speakers are randomly sampled from all $N$ training speakers. However, there are only 588 in-domain Farsi speakers out of a total of 9077 training speakers. This bias possibly leads to speaker embeddings that are sub-optimal towards the target-domain. A common transfer learning technique is to fine-tune a pre-trained model on the target-domain data with the goal to correct the data distribution mismatch between the training and target-domain. Due to the tendency of neural networks to easily overfit on small datasets, we opt to learn a robust embedding that performs reasonably well on both the available out-of-domain VoxCeleb data and target-domain DeepMine training data.

We correct the bias towards the VoxCeleb and LibriSpeech corpus by equalizing the sample probability for each domain. During the construction of the batches, subsequent selections of the $S$ speakers cover a set of all 588 Farsi speakers and 588 random speakers from both the VoxCeleb and LibriSpeech domain. When the set runs empty, the similarity matrix $\pmb{S}$ is updated and 588 new speakers are randomly selected from the out-of-domain data to allow reiteration of the batch generation process. This process assigns more importance towards samples from hard speakers in the target-domain, while still allowing the network to learn from samples of challenging out-of-domain speakers.




\subsection{Adaptive s-normalization with language offset}
\label{sec:ld_s_norm}

Based on \cite{s_norm}, we set the imposter cohort of the adaptive s-normalization to contain in-domain Farsi data only. However, an unknown portion of the test utterances in the SdSVC trials is English.
In case of a speaker verification trial with language mismatch, this will result in an overestimated mean imposter score for the Farsi enrollment model, as it will only be compared against Farsi imposters. We introduce a language-dependent offset in the adaptive s-normalization procedure to compensate for this effect.

Given a trial score $ s(e, t) $ between the enrollment model $e$ and test utterance $t$, the  language-dependent s-normalized score is defined as:
\begin{equation}
\label{s_norm_offset}
s(e, t)_{n} = \frac{s(e, t) - \mu(S_t) }{\sigma(S_t)} + \frac{s(e, t) - (\mu(S_e) - \alpha) }{\sigma(S_e)}.
\end{equation}
with $S_i$ the set of scores of the speaker embedding $i$ against its top-N imposter cohort, with $\mu(S_i)$ the mean of those scores and $\sigma(S_i)$ the standard deviation. $\alpha$ is the language-dependent compensation offset. It is defined as zero if there is no language mismatch detected and in that case regular adaptive s-norm is applied. When during test time the test utterance is detected to be English, we enable the language offset. Given $\mu_{S_{FA}}$ as the expected mean imposter score of Farsi imposters against a Farsi speaker and $\mu_{S_{USA}}$ as the expected mean imposter score of USA-English imposters against a Farsi speaker, we define this compensation offset $\alpha$ as $\mu_{S_{FA}} - \mu_{S_{USA}}$. The mean imposter values can be easily estimated on the speaker prototypes stored in the AAM-softmax module by applying s-norm on the relevant prototypes.

To detect the language of the test utterance given its embedding, we train a Language Identification (LID) module based on a Gaussian Backend (GB) \cite{gb} modeled on the $L_2$-normalized AAM speaker prototypes of the Farsi and the USA speakers. However, there will be a mismatch between the English spoken by a native Farsi speaker and a USA citizen. To compensate for this effect we interpolate between the GB mean vector for the USA language class $\pmb{\mu}_{USA}$ and the mean vector corresponding with Farsi $\pmb{\mu}_{FA}$ and set the expected mean embedding of the English model to $0.75\pmb{\mu}_{USA} + 0.25\pmb{\mu}_{FA}$. This adapted language model should be able to robustly detect English spoken by a native Farsi speaker.



\subsection{Final submission}

The IDLab final submission for the SdSVC consists of a fusion of the five proposed ECAPA-TDNN systems fine-tuned with domain-balanced HPM combined with language-dependent s-normalization with the LID labels extracted from System 1. The fusion is realized on the score level by taking a weighted average over the calibrated scores of each individual system. The systems that incorporate Res2 modules are given double the weight in the averaging compared to the other systems.



\section{Results and additional experiments}
\label{sec:exp}


\subsection{ECAPA-TDNN baseline performance}

The baseline performance of the ECAPA-TDNN architectures on the SdSVC evaluation data is shown in Table~\ref{tab:exp_results}. We also keep track of results on the original VoxCeleb1 test set to verify the system is not overfitting on the training data. No s-normalization is used for the VoxCeleb1 evaluation results.

These baseline single system implementations show strong and similar performance on both the SdSVC and VoxCeleb data, reaching up to an EER and MinDCF of 2.13\% and 0.0938 respectively on the SdSVC test set. System~$4$ with SE-Res2Blocks and summed inputs slightly outperforms the other equally sized systems, while its much larger counterpart System~$5$ only delivers a small performance gain.

\subsection{Domain-balanced HPM fine-tuning}

The impact of domain-balanced HPM fine-tuning on the baseline systems can be found in Table \ref{tab:exp_results}. After fine-tuning, all systems perform significantly better on the SdSVC test set with an average improvement of 24.1\% in EER and 21.8\% in MinDCF. The performance difference between System~4 and System~5 has vanished on the SdSVC test set. Notably, results on the VoxCeleb1 test set remain strong and often improve after applying domain-balanced HPM, despite the reduced VoxCeleb sampling frequency. 

We conduct additional experiments to separately study the impact of the increased sampling frequency of Farsi and the focus on harder samples during training. Results of these experiments can be found in Table~\ref{tab:hpm_effects}. We fine-tune the System~5 baseline with the protocol described in Section~\ref{sec:hpm}, but do not take the speaker similarity into account and just randomly sample imposter speakers from the same domain. One experiment balances the domain of speakers (\textit{balanced}) while another experiment exclusively samples from the in-domain (\textit{Farsi}) training set. In addition, we compare our domain-balanced HPM approach against the broad HPM of Section~\ref{sec:bhpm} and against an HPM variant that only samples from Farsi speakers.

\begin{table}[h]
  \caption{Effects of fine-tuning (FT) and our proposed HPM strategies.}
  \label{tab:hpm_effects}
  \centering
  \begin{tabular}{ccccc}
    \toprule
    \multicolumn{1}{c}{\textbf{Method}} & \multicolumn{1}{c}{\textbf{Domain}} & \multicolumn{1}{c}{\textbf{Vox1}} & \multicolumn{2}{c}{\textbf{SdSVC 2020}} \\
    \cmidrule(lr){3-3} \cmidrule(lr){4-5}
    \multicolumn{2}{c}{} & \multicolumn{1}{c}{\textbf{EER(\%)}} &  \multicolumn{1}{c}{\textbf{EER(\%)}} & \multicolumn{1}{c}{\textbf{MinDCF}} \\
    \midrule
    baseline & - & 0.87 & 2.13 & 0.0938 \\
    \midrule
    \midrule
    HPM & Farsi & 2.00 & 2.01 & 0.0910 \\
    HPM & broad & 0.83 & 1.98 & 0.0875 \\
    HPM & balanced & \textbf{0.79} & \textbf{1.69} & \textbf{0.0759} \\
    \midrule
    FT & Farsi & 6.05 & 1.83 & 0.0854 \\
    FT & balanced & 0.87 & 1.82 & 0.0802 \\
    \bottomrule
  \end{tabular}
\end{table}

Basic fine-tuning of the systems on SdSVC training data only, increases the in-domain performance significantly with a relative improvement of 14.1\% and 9.0\% in EER and MinDCF respectively. Balancing the sampling frequency however, prevents the degradation on the VoxCeleb1 test set and further improves the MinDCF by 6.1\% relative. The EER remains stable. This indicates that it is worthwhile to keep out-of-domain performance stable while fine-tuning the systems.

The importance of domain-balancing increases when applying our proposed HPM strategy. As the balance between the domain sampling increases, so does the performance on both evaluation sets. Incorporating HPM on top of domain-balanced sampling shows to be beneficial and increases relative performance with 7.1\% and 5.4\% in EER and MinDCF respectively.




\subsection{Language-dependent score normalization}

As shown in Table~\ref{tab:exp_results}, the language-dependent variant of our adaptive s-normalization system further improves EER and MinDCF values on average with 3.3\% and 2.3\% respectively on the SdSVC test set. While modest, the improvement is consistent and easy applicable in the scoring backend.

To analyze the impact of different imposter speaker cohorts, we analyze the HPM domain-balanced System~5 with different s-norm configurations. The results on the SdSVC test set are provided in Table~\ref{tab:s_norm_effects}. The imposter cohort is restricted to the top-40 most similar imposters for all experiments.

\begin{table}[h]
  \caption{Effects of cohort selection in s-normalization.}
  \label{tab:s_norm_effects}
  \centering
  \begin{tabular}{cccc}
    \toprule
    \multicolumn{1}{c}{\textbf{Imposter Cohort}} & \multicolumn{1}{c}{\textbf{EER(\%)}} & \multicolumn{1}{c}{\textbf{MinDCF}} \\
    \midrule
    no s-normalization & 2.14 & 0.0947 \\
    \midrule
    \midrule
    VoxCeleb & 2.46 & 0.1303 \\
    Farsi & \textbf{1.69} & \textbf{0.0759} \\
    VoxCeleb + Farsi & 1.72 & 0.0762 \\
    \bottomrule
  \end{tabular}
\end{table}

The results clearly illustrate that a cohort restricted to the available in-domain training data proves to be the most optimal configuration. We notice a relative improvement of 21\% and 19.9\% in EER and MinDCF respectively over a system without s-norm.

\subsection{Final submission}

The final score-based fusion of the single systems fine-tuned with domain-balanced HPM and language-dependent score normalization  results in an EER of 1.45\% and a MinDCF of 0.0651 as shown in Table \ref{tab:exp_results}. Fusion of all systems leads to a relative improvement over System~5 of 11\% and 11.9\% in EER and MinDCF respectively on the SdSVC test set. This shows that minor architectural variations can prove sufficient to learn complementary speaker embeddings.

\section{Conclusion}
\label{sec:conclusion}

In this paper we presented HPM as a computationally efficient hard negative mining strategy to fine-tune a speaker embedding extractor towards out-of-domain Farsi data. Furthermore, a correct configuration of s-normalization has proved to be crucial to handle the cross-lingual trials presented in the SdSV Challenge 2020. A fusion of five systems based on our ECAPA-TDNN architecture in conjunction with the proposed techniques resulted in a final top-scoring submission on Task 2 of the SdSVC with an EER of 1.45\% and a MinDCF of 0.065.

\bibliographystyle{IEEEtran}

\bibliography{mybib}

\end{document}